{}%
\documentclass[twocolumn,showpacs,preprintnumbers,amsmath,amssymb,superscriptaddress]{revtex4}

\usepackage{graphicx}
\usepackage{dcolumn}
\usepackage{bm}

\newcommand{\al}{$\alpha$}
\newcommand{\raa}{($\alpha$,$\alpha$)}

\begin{document}

\preprint{APS/123-QED}

\title{High precision $^{89}$Y($\alpha$,$\alpha$)$^{89}$Y scattering at
  low energies
}

\author{G. G.\,Kiss}%
\email{ggkiss@atomki.hu}
\altaffiliation [present address:]{ Laboratori Nazionali del Sud, INFN, Catania, Italia}
\affiliation{%
Institute of Nuclear Research (ATOMKI), H-4001 Debrecen, Hungary}%
\author{P.\,Mohr}%
\affiliation{%
Diakonie-Klinikum, D-74523 Schw\"abisch Hall, Germany}%
\author{Zs.\,F\"ul\"op}%
\affiliation{%
Institute of Nuclear Research (ATOMKI), H-4001 Debrecen, Hungary}%
\author{D.\,Galaviz}%
\altaffiliation [present address:]{ Centro de F\'isica Nuclear da Universidade de Lisboa, 1649-003, Lisbon, Portugal}
\affiliation{ Instituto de Estructura de la Materia, CSIC, E-28006 Madrid, Spain}
\author{Gy.\,Gy\"urky}%
\author{Z. Elekes}%
\author{E.\,Somorjai}%
\affiliation{%
Institute of Nuclear Research (ATOMKI), H-4001 Debrecen, Hungary}%
\author{A.\,Kretschmer}%
\author{K.\,Sonnabend}%
\affiliation{%
Institut f\"ur Kernphysik, Technische Universit\"at Darmstadt, D-64289 Darmstadt, Germany}%
\author{A.\,Zilges}%
\affiliation{%
Institut f\"ur Kernphysik, Universit\"at zu K\"oln, D-50937 K\"oln, Germany}
\author{M.\,Avrigeanu}%
\affiliation{%
"Horia Hulubei" National Institute for Physics and Nuclear Engineering, 76900 Bucharest, Romania}
\date{\today}

\begin{abstract}
Elastic scattering cross sections of the $^{89}$Y($\alpha$,$\alpha$)$^{89}$Y
reaction have been measured at energies E$_{c.m.}$ = 15.51 and
18.63 MeV. The high precision data for the semi-magic $N = 50$ nucleus
$^{89}$Y are used to derive a local potential and to evaluate the predictions
of global and regional $\alpha$-nucleus potentials. The variation of the
elastic alpha scattering cross sections along the $N = 50$ isotonic chain is
investigated by a study of the ratios of angular distributions for
$^{89}$Y($\alpha$,$\alpha$)$^{89}$Y and $^{92}$Mo($\alpha$,$\alpha$)$^{92}$Mo
at E$_{c.m.} \approx$ 15.51 and 18.63 MeV. This ratio is a very
sensitive probe at energies close to the Coulomb barrier, where scattering
data alone is usually not enough to characterize the different
potentials. Furthermore, $\alpha$-cluster states in $^{93}$Nb = $^{89}$Y
$\otimes$ $\alpha$ are investigated.
\end{abstract}


\maketitle

\section{Introduction}

Alpha-nucleus potentials are basic ingredients for the calculation of reaction
cross sections with $\alpha$ particles in the entrance or exit channel. These
reaction cross sections are included in the calculation of stellar reaction rates
in nuclear astrophysics which have to be determined at stellar temperatures
corresponding typically to sub-Coulomb energies for reactions involving
$\alpha$ particles.

In several astrophysical applications --- such as modeling
the nucleosynthesis in explosive scenarios like $p$ process --- the
reaction rates are taken from statistical model calculations
\cite{rau01, arn03}. These calculations utilize global
alpha-nucleus optical potential parameter sets. Considerable efforts
have been devoted in recent years to improve the knowledge of the
alpha-nucleus optical potential \cite{avr03, avr07, kum06, dem02}. The
extensive use of the statistical 
model calculations requires further experimental tests for the
global parameterizations.

The optical potential combines a Coulomb term with a complex nuclear
potential, composed of real and imaginary parts. The variation of the
potential parameters of the real part as a function of mass and energy is
smooth and relatively well understood \cite{atz96}. On the contrary, the
imaginary part of the optical potential is strongly energy-dependent
especially at energies around the Coulomb barrier. Therefore, tests of global
$\alpha$-nucleus potentials need to focus on experimental information at
energies as close as possible to the astrophysically relevant energy
region. One possible way of testing the different potential parameterizations
is to carry out alpha elastic scattering experiments and compare the measured
angular distributions to the corresponding predictions from the global
potential parameterizations. However, at astrophysical energies the
alpha-nucleus elastic scattering cross section is non-diffractive and
dominated by the Rutherford component. Therefore, the experiments have to be
carried out at slightly higher energies with high precision. From the analysis
of the measured angular distributions the parameters of the potential can be
derived and have to be extrapolated down to the astrophysically relevant energy region where the relevant alpha-particle induced-reactions are taking place. 

The present study focuses on the determination of optical potentials from
elastic scattering cross sections. It is an interesting and still open
question whether reaction cross sections can be predicted from such a
potential that has only been adjusted to scattering data. But it remains as a
final goal to find a perfect potential that is able to reproduce elastic
scattering and reaction cross sections and furthermore properties of cluster
states like excitation energies and decay widths.

Several alpha elastic scattering experiments on even-even nuclei $^{92}$Mo,
$^{106}$Cd, $^{112,124}$Sn, and $^{144}$Sm have been performed at ATOMKI in
recent years \cite{ful01, kis06, gal05, moh97}, in order to investigate the
behavior of alpha-nucleus optical potentials. This work presents the elastic
scattering experiment performed on the $^{89}$Y nucleus to study further the
systematic behavior of the optical potentials at low energies.  In all of
these cases complete angular distributions have been measured at energies
close to the Coulomb barrier. The chosen energies were low enough to be close
to the region of astrophysical interest and high enough that the scattering
cross section differs sufficiently from the Rutherford one.

Previous studies have focused on semi-magic even-even nuclei with $N = 82$
($^{144}$Sm), $N = 50$ ($^{92}$Mo), and $Z = 50$ ($^{112,124}$Sn). As a
natural extension of previous work, we are presently working on nuclei that
are either non-magic or even-odd nuclei. Thus, a main motivation for our
latest experiment on the non-magic nucleus $^{106}$Cd was to analyze the
influence of shell closures on the $\alpha$-nucleus potential \cite{kis06}. As
a continuation of the systematic study of the behavior of the optical
potentials, this work presents the elastic scattering experiment performed on
the neutron-magic ($N = 50$), proton-odd ($Z = 39$) nucleus $^{89}$Y. Angular
distributions have been measured at E$_{c.m.}$= 15.51 and 18.63 MeV, just
above the Coulomb barrier (the height of the Coulomb barrier for the $^{89}$Y
$\otimes$ $\alpha$ system is about 15 MeV). At these energies a reliable test
for the global parametrization is possible using the new high precision
data. Furthermore, we can use data from literature \cite{wit75,eng82} to
investigate the variation of the imaginary part of the optical potential
between E$_{c.m.}$ $\approx$ 15.5 and 23.9 MeV in small steps of approximately
2 MeV. In addition, experimental angular distributions are available at higher
energies \cite{Bri72,Bin69,Als66,Ber79}. A local optical potential for
$^{89}$Y can be established from these data over a wide range of
energies. This was a further motivation to study \al\ scattering on $^{89}$Y. 

A global alpha-nucleus optical potential must be able to provide a correct
prediction for the elastic scattering cross section, and to describe the
variation of the angular distributions along isotopic and isotonic
chains. This is especially important for the extrapolation to unstable nuclei
where the potential cannot be derived from experimental scattering data.
Galaviz \emph{et al.} \cite{gal05} measured the elastic scattering cross
sections of the $^{112,124}$Sn($\alpha$,$\alpha$)$^{112,124}$Sn reactions at
E$_{\alpha}$ = 19.5 MeV over a broad angular range with small uncertainties of
about $\approx$ 3-4\%. The study of both the proton- and neutron-rich stable
tin isotopes provided important information about the changes of the potential
parameters with the neutron number along the $Z = 50$ isotopic chain. The
ratio of the measured cross sections showed an oscillation pattern which was
very pronounced in the backward angle region. The analysis of this ratio
provides a further constraint for global \al -nucleus potentials, in
particular for the extrapolation to unstable nuclei. It needs to be
highlighted that all global \al -nucleus optical potentials failed to
reproduce either the amplitude or the phase of this oscillation pattern in the
ratio of the measured angular distributions of $^{112}$Sn and $^{124}$Sn at
backward angles \cite{gal05}. This oscillation feature of the experimental
data was first observed using our precise data on $^{112}$Sn and $^{124}$Sn,
but it was hidden previously because the typical uncertainties of the elastic
scattering cross sections were of the order of 10-15\% \cite{avr_ad}.

As an extension of our previous work on the isotopic chain at $Z =
50$, now we investigate the behavior of the optical potential
parameters along the $N=50$ isotonic chain. We have measured
$\alpha$ scattering data of $^{89}$Y with small uncertainties over the whole
angular range. This is a prerequisite to study the ratio of the 
Rutherford normalized cross sections of
$^{89}$Y($\alpha$,$\alpha$)$^{89}$Y from this work and
$^{92}$Mo($\alpha$,$\alpha$)$^{92}$Mo previously studied by F\"ul\"op
\emph{et al.} \cite{ful01} at E$_{c.m.}$ = 13.20,
15.69 and 18.62 MeV. Further studies could be done on the $N=50$ nuclei
$^{86}$Kr, $^{87}$Rb, and $^{88}$Sr. Low-energy data for the $N=50$ nucleus
$^{90}$Zr at 15\,MeV \cite{Wat71} were already analyzed in \cite{atz96}.

This paper is organized as follows. In Sec.~\ref{sec:exp} we describe our
experimental procedure. In a first analysis a local optical potential is
derived from the new experimental data and the available scattering data in
literature, and furthermore bound state properties of cluster states in the
nucleus $^{93}$Nb = $^{89}$Y $\otimes$ $\alpha$ are studied
(Sec.~\ref{sec:local}).
The measured angular distributions as well as the
ratio of the elastic scattering cross sections of the two $N=50$ nuclei
$^{89}$Y and $^{92}$Mo are compared to predictions using several global
optical potential parameterizations in Sec.~\ref{sec:global}; additionally,
calculations are compared to angular distributions at higher energies and to
excitation functions from literature \cite{wit75,eng82}. Finally, conclusions
are drawn in Sec.~\ref{sec:sum}.

\section{Experimental technique}
\label{sec:exp}

The experiment was carried out at the cyclotron laboratory of ATOMKI,
Debrecen. A similar experimental setup was used also in the previous
experiments \cite{ful01, gal05, moh97, kis06} and is described
in more detail in \cite{mat89}. The following paragraphs provide a
short description of the experimental procedure.
 
\subsection{Targets and scattering chamber}

The targets were produced by evaporation of metallic yttrium onto thin carbon
foils ($\approx$ 20 $\mu$g/cm$^{2}$). The target thickness was approximately 200
$\mu$g/cm$^{2}$. The targets were mounted on a remotely controlled target
ladder in the center of the scattering chamber. The stability of the targets
was checked continuously with monitor detectors (see below) during the
experiment. 

The E$_{lab}$ = 16.21 and 19.47 MeV energy alpha beam (beam current: 150 pnA)
was lead to the scattering chamber through an analyzing magnet. Since the
energy stability of the beam is important for the experiment, the widths of
the slits at the entrance and the exit of the magnet were small (1 mm).
Moreover, the beam energy was monitored during the experiment with the monitor
detectors (see below). The total uncertainty of the beam energy was found to
be less than $ 0.5 \%$.

An aperture of 2 x 6 mm was mounted on the target ladder
to check the beam position and size of the beamspot before and after every
change of the beam energy or current. We optimized the beam until not more
than 1\% of the total beam current could be measured on this aperture. As a
result of the procedure, the horizontal size of the beamspot was below
2 mm during the whole experiment which is crucial for the precise
determination of the scattering angle. 

\subsection{Detectors and data acquisition}

Four ion implanted silicon detectors with active areas of 50 mm$^2$ were used for the measurement of the angular distributions. The detectors were mounted in pairs separated by 10$^\circ$. The solid angles covered by the two detector pairs were $\Delta$$\Omega$=1.56 x 10$^{-4}$ and $\Delta$$\Omega$=1.81 x 10$^{-4}$. The ratios of solid angles of the different detectors were checked  by measurements at overlapping angles with good statistics.
 
\begin{figure*}[htb]
\resizebox{1.5\columnwidth}{!}{\rotatebox{270}{\includegraphics[clip=]{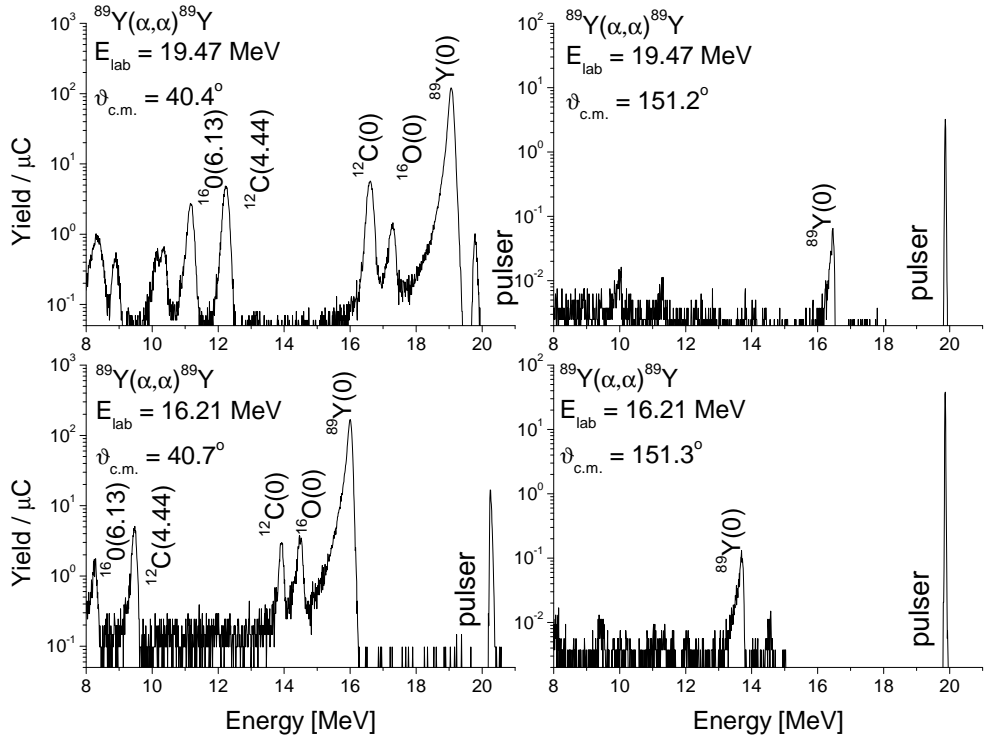}}}
\caption{ 
\label{fig:spec}
Typical spectra at E$_{c.m.}$ = 15.51 MeV and 18.63 MeV,
$\vartheta$$_{c.m.}$ $\approx$ 40$^\circ$ and 150$^\circ$. The peak
from elastic $^{89}$Y-$\alpha$ scattering is well resolved from both
the $^{12}$C-$\alpha$ and $^{16}$O-$\alpha$ elastic
scattering. The pulser peak used for the dead time
correction is also shown. Note the logarithmic scale of all spectra.}
\end{figure*}
 
In addition, two detectors were mounted at a larger distance on the wall of
the scattering chamber at fixed angles $\vartheta$=$\pm$15$^\circ$ left and
right to the beam axis. These detectors were used as monitor detectors during
the experiment to normalize the measured angular distribution and to determine
the precise position of the beam on the target. The solid angle of these
detectors was $\Delta$$\Omega$=8.2 x 10$^{-6}$.

The signals from all detectors were processed using charge-sensitive
preamplifiers. The output signals were further amplified by main amplifiers
and fed into analog-to-digital converters. Since the elastic scattering cross
sections at forward angles differ several orders of magnitude from the
one measured at backward angles, a reliable dead time correction is
crucial. The data were collected using the WinTMCA system which provides an
automatic dead time control. This automatically determined dead time was
verified using a pulser in all spectra. 

The energy of the first of the excited state of the $^{89}$Y nucleus is 908.97 keV
\cite{nndc}. There is a large difference between the spin of the ground and
the first excited states (1/2$^-$ and 9/2$^+$ respectively). Therefore the
expected inelastic scattering cross section is very low (below 10$^{-4}$
mbarn, calculated with the DWUCK code \cite{dwuck}) at the measured energies. 

This fact explains why events corresponding to inelastic alpha scattering on $^{89}$Y are missing from the spectra. 
Typical spectra are shown in Fig.~\ref{fig:spec}. 
The relevant peaks
from elastic $^{89}$Y-$\alpha$ scattering are well separated from
elastic and inelastic peaks of target contaminations, and -- as expected --
peaks from inelastic $\alpha$ scattering on $^{89}$Y are not
visible. 

\subsection{Angular calibration}

Knowledge on the exact angular position of the detectors is of crucial importance 
for the precision of a scattering experiment since the Rutherford cross section
depends sensitively on the angle. The uncertainty in the angular distribution
is dominated by the error of the scattering angles in the forward
region. A tiny uncertainty of $\Delta$$\vartheta$ = 0.3$^\circ$ results in a
significant error of approximately 5\% in the Rutherford normalized cross sections at very forward angles.

\begin{figure}
\resizebox{0.8\columnwidth}{!}{\rotatebox{270}{\includegraphics[clip=]{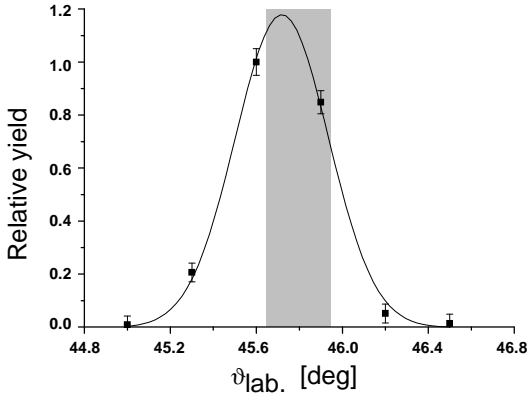}}}
\caption{
\label{fig:angcalib}
Relative yield of $^{12}$C recoil nuclei in coincidence with elastically scattered $\alpha$ particles. The shaded area represents the angle and uncertainties expected from the reaction kinematics. A Gaussian curve fitted to the experimental data is shown to guide the eye.} 
\end{figure}

To determine the scattering angle precisely, we measured kinematic
coincidences between elastically scattered alpha particles and the
corresponding $^{12}$C recoil nuclei using a pure carbon foil target.  One
detector was placed at $\vartheta$ = 70$^\circ$, and the signals from the
elastically scattered alpha particles on $^{12}$C were selected as gates for
the other detector which moved around the expected $^{12}$C recoil angle
$\vartheta$ = 45.85$^\circ$. 
Figure~\ref{fig:angcalib} shows the relative
yield of the $^{12}$C recoil nuclei in coincidence with elastically scattered
alpha particles as a function of the $^{12}$C recoil angle. The final angular
uncertainty was found to be $\Delta$$\vartheta$ $\leq$ 0.15$^\circ$.

\subsection{Elastic scattering cross sections and data analysis}

Complete angular distributions between 20$^{\circ}$ and 170$^{\circ}$
were measured at energies of E$_{\alpha}$ = 16.21 and 19.47 MeV in
1$^\circ$ (20$^\circ$ $\leq$ $\vartheta$ $\leq$ 100$^\circ$),
1.5$^\circ$ (100$^\circ$ $\leq$ $\vartheta$ $\leq$ 140$^\circ$) and
2$^\circ$ (140$^\circ$ $\leq$ $\vartheta$ $\leq$ 170$^\circ$) steps.

The statistical uncertainties varied between 0.1\% (forward angles) and 4\% (backward angles). The count rates \textit{N($\vartheta$)} have been normalized to the yield of the monitor detectors \textit{N$_{Mon.}$($\vartheta$=15$^\circ$)}:

\begin{equation}
\left(\frac{d\sigma}{d\Omega}\right)(\vartheta)\,=\left(\frac{d\sigma}{d\Omega}\right)_{Mon.}\frac{N(\vartheta)}{N_{Mon.}}\frac{\Delta\Omega_{Mon.}}{\Delta\Omega},
\end{equation}
with $\Delta$$\Omega$ being the solid angles of the detectors. The cross section at the position of the monitor detectors is taken as pure Rutherford. The relative measurement eliminates the typical uncertainties of absolute measurements, coming mainly from changes in target and from the beam current integration. 

The measured angular distributions are shown in Fig.~\ref{fig:ruth}. The
lines are the result of optical model predictions using global
$\alpha$-nucleus potentials (see
Sec.~\ref{sec:global}). The measured absolute cross sections cover
five orders of magnitude between the highest (forward angles at
E$_{\alpha}$=16.21 MeV) and the lowest cross sections (backward angle
at E$_{\alpha}$=19.47 MeV) with almost the same accuracy (4-5\% total
uncertainty). This error is mainly caused by the uncertainty of the
determination of the scattering angle in the forward region and from the
statistical uncertainty in the backward region.

\section{Local optical potential and $\alpha$-cluster states}
\label{sec:local}

\subsection{Local folding potential}
\label{sec:fold}

The complex optical model potential (OMP) is given by:

\begin{equation}
U(r)\,=V_{C}(r)+V(r)+iW(r),
\end{equation}

where \textit{V$_C$(r)} is the Coulomb potential, \textit{$V(r)$}, and
\textit{$W(r)$} are the real and the imaginary parts of the nuclear potential,
respectively. The real part of the potential is calculated from the folding
procedure \cite{Kob84,Sat79} using a density-dependent nucleon-nucleon
interaction. The calculated folding potential is adjusted to the experimental
scattering data by two parameters

\begin{equation}
V(r) = \lambda \, V_{F}(r/w)
\label{eq:fold}
\end{equation}

where $\lambda \approx 1.1 - 1.4$ is the potential strength parameter
\cite{atz96} and $w \approx 1.0 \pm 0.04$ is the width parameter that
slightly modifies the potential width. (Larger deviations of the width
parameter $w$ from unity would indicate a failure of the folding potential.)
The nuclear densities of $^{89}$Y and $\alpha$ are derived from the measured
charge density distributions \cite{Vri87}.
For details of the folding potential see also \cite{Abe93,Mohr97}.

The imaginary part $W(r)$ is taken in the usual Woods-Saxon
parametrization. For the fits to the experimental data we use a sum of
volume and surface potential:

\begin{equation}
W(r) = W_V \times f(x_V) + 4 \, W_S \times \frac{df(x_S)}{dx_S}
\end{equation}

with the potential depths $W_V$ and $W_S$ of the volume and surface parts and 

\begin{equation}
f(x_i) = \frac{1}{1+\exp{(x_i)}}
\end{equation}
and $x_i = (r-R_i*A_T^{1/3})/a_i$ with the radius parameters $R_i$, the
diffuseness parameters $a_i$, and $i=S,V$.

\begin{figure}
\includegraphics[bbllx=15,bblly=30,bburx=440,bbury=375,width=\columnwidth,clip=]{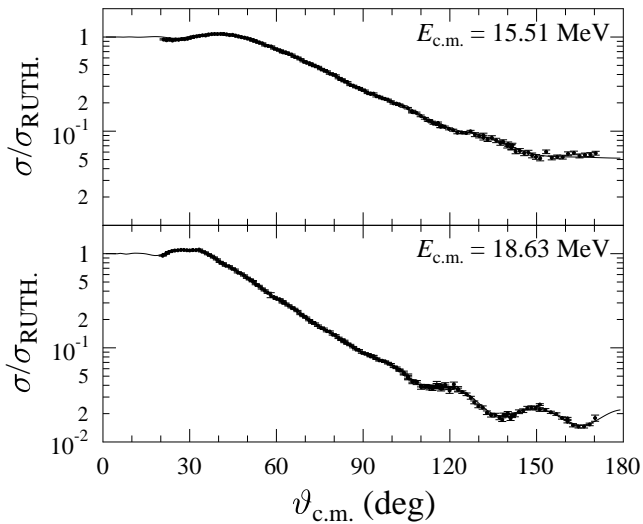}
\caption{
\label{fig:scatlow}
Rutherford normalized elastic scattering cross sections of
$^{89}$Y($\alpha,\alpha$)$^{89}$Y reaction at E$_{c.m.}$ = 15.51 and 18.63 MeV
versus the angle in center-of-mass frame. The lines are the results from the
local potential which is fitted to the experimental data. The parameters of
the fits are listed in Table \ref{tab:pot}.
}
\end{figure}

The adjustment of the OMP parameters leads to an excellent description of the
new experimental data (see Fig.~\ref{fig:scatlow}). The parameters of the
potentials are listed in Table \ref{tab:pot}. 

\begin{table*}
  \caption{\label{tab:pot}
    Parameters of the local potentials that were derived from elastic
    scattering angular distributions in a wide energy range (see
    Figs.~\ref{fig:scatlow} and \ref{fig:scatall}).
}
\begin{center}
\begin{tabular}{cccccccccccccc}
\multicolumn{1}{c}{$E_{\rm{c.m.}}$} 
& \multicolumn{1}{c}{$\lambda$} 
& \multicolumn{1}{c}{$w$} 
& \multicolumn{1}{c}{$J_R$} 
& \multicolumn{1}{c}{$r_{R,rms}$} 
& \multicolumn{1}{c}{$J_I$} 
& \multicolumn{1}{c}{$r_{I,rms}$} 
& \multicolumn{1}{c}{$W_{V0}$} 
& \multicolumn{1}{c}{$R_V$} 
& \multicolumn{1}{c}{$a_V$} 
& \multicolumn{1}{c}{$4 \times W_{S0}$} 
& \multicolumn{1}{c}{$R_S$} 
& \multicolumn{1}{c}{$a_S$}
& experimental \\
\multicolumn{1}{c}{(MeV)} 
& & 
& \multicolumn{1}{c}{(MeV\,fm$^3$)} 
& \multicolumn{1}{c}{(fm)} 
& \multicolumn{1}{c}{(MeV\,fm$^3$)} 
& \multicolumn{1}{c}{(fm)} 
& \multicolumn{1}{c}{(MeV)}
& \multicolumn{1}{c}{(fm)} 
& \multicolumn{1}{c}{(fm)} 
& \multicolumn{1}{c}{(MeV)}
& \multicolumn{1}{c}{(fm)} 
& \multicolumn{1}{c}{(fm)} 
& data from \\
\hline
15.5  & 1.350  & 0.986  & 342.3  & 4.905  & 49.9  & 5.953  & -11.3  & 1.656  & 0.486  & -8.3
& 1.350  & 0.548 & this work \\
18.6  & 1.353  & 0.979  & 335.3  & 4.868  & 46.9  & 4.975  & -20.0  & 1.331  & 0.713  & -25.8
& 1.487  & 0.255 & this work \\
20.1  & 1.328  & 0.985  & 334.9  & 4.897  & 45.0  & 5.262  & -22.3  & 1.197  & 0.982  & -55.7
& 1.501  & 0.174 & Ref.~\cite{wit75} \\
22.0  & 1.352  & 0.969  & 324.7  & 4.817  & 47.2  & 5.856  & -16.9  & 1.585  & 0.822  & -42.7
& 1.460  & 0.502 & Ref.~\cite{wit75} \\
23.9   & 1.315  & 0.989  & 335.8  & 4.917  & 46.6  & 4.955  & -21.9  & 1.656 &
0.590  & -71.2  & 1.435  & 0.605 & Ref.~\cite{wit75} \\ 
23.9  & 1.326  & 0.984  & 334.5  & 4.896  & 44.7  & 4.735  & -23.6  & 1.672  &
0.610  & -78.5 & 1.423  & 0.678 & Ref.~\cite{eng82} \\ 
40.2  & 1.268  & 1.002  & 328.0  & 4.988  & 57.5  & 5.672  & -18.8  & 1.492  & 0.799  & -26.3
& 1.475  & 0.429 & Ref.~\cite{Als66} \\
62.2  & 1.296  & 1.007  & 328.3  & 5.016  & 65.6  & 5.835  & -19.5  & 1.634  & 0.544  & -56.7
& 1.388  & 0.365 & Ref.~\cite{Bin69} \\
158.9 & 1.317  & 0.985  & 260.0  & 4.930  & 82.3  & 5.785  & -22.1  & 1.457  & 0.757  & $-$
& $-$  & $-$ & Ref.~\cite{Bri72} \\
\hline
\end{tabular}
\end{center}
\end{table*}

It has to be noted that the folding potential in the real part may be replaced
by a Woods-Saxon potential. A similar quality of the fits can be obtained in
this case. However, the adjustment of three Woods-Saxon parameters in the real
part and six parameters in the imaginary part leads to a variety of parameter
sets with comparable $\chi^2$. This problem has been reduced by \cite{avr07}
where a microscopic Woods-Saxon potential was derived from a folding
procedure. This potential is used as the base of one of the global potentials
studied in Sect.~\ref{sec:global}.

The above calculation with folding potentials has been repeated for the
$^{89}$Y($\alpha$,$\alpha$)$^{89}$Y scattering data available in literature
from about 20 MeV up to about 170 MeV. The analyzed data have been measured by
Brissaud {\it et al.}~\cite{Bri72} at 166 MeV ($E_{\rm{c.m.}} = 158.9$ MeV),
Bingham {\it et al.}~\cite{Bin69} at 65 MeV ($E_{\rm{c.m.}} = 62.2$ MeV),
Alster {\it et al.}~\cite{Als66} at 42 MeV ($E_{\rm{c.m.}} = 40.2$ MeV),
England {\it et al.}~\cite{eng82} at 25 MeV ($E_{\rm{c.m.}} = 23.9$ MeV), and
Wit {\it et al.}~\cite{wit75} at 25, 23, and 21 MeV ($E_{\rm{c.m.}} = 23.9$,
22.0, and 20.1 MeV) with an additional excitation function at backward
angles. The data measured by Berinde {\it et al.} \cite{Ber79} at 27.3 MeV
($E_{\rm{c.m.}} = 26.1$ MeV) were excluded because all fits were of poor
quality and required width parameters $w$ deviating strongly from unity. The
new and challenging technique for the measurement of angular distributions
using a position-sensitive detector in the work of Berinde {\it et
  al.}~\cite{Ber79} was unfortunately not further developed; this may indicate
that there are problems with the data of \cite{Ber79}.

At 25 MeV two data sets are available. The data by England {\it et
  al.}~\cite{eng82} cover the whole angular range from forward to backward
directions. As expected, at forward angles the data agree with the Rutherford
cross section, and thus the uncertainty of the absolute normalization of the
data is of the order of a few per cent. The data by Wit {\it et
  al.}~\cite{wit75} focus on the backward region and start around
$\vartheta_{\rm{c.m.}} \approx 50^\circ$ where the cross section is about
$20\,\% - 30\,\%$ of the Rutherford cross section. Although an uncertainty of
about $\pm 15\,\%$ is claimed in \cite{wit75}, the comparison to the data by
England {\it et al.}~\cite{eng82} shows that the data by Wit {\it et al.}~have
to be reduced by a factor of 1.45 to come into agreement with the England {\it
  et al.}~data. The origin of this discrepancy remains unclear. Because
typical normalization problems (e.g.\ uncertainties of the target thickness)
apply probably to all measurements of the $^{89}$Y($\alpha$,$\alpha$)$^{89}$Y
cross section in \cite{wit75}, we use the reduction factor of 1.45 for all
data for $^{89}$Y measured by Wit {\it et al.}~\cite{wit75}, i.e.\ the three
angular distributions at 21, 23, and 25 MeV and the excitation function at
backward angles.

\begin{figure}
\includegraphics[bbllx=15,bblly=30,bburx=440,bbury=665,width=\columnwidth,clip=]{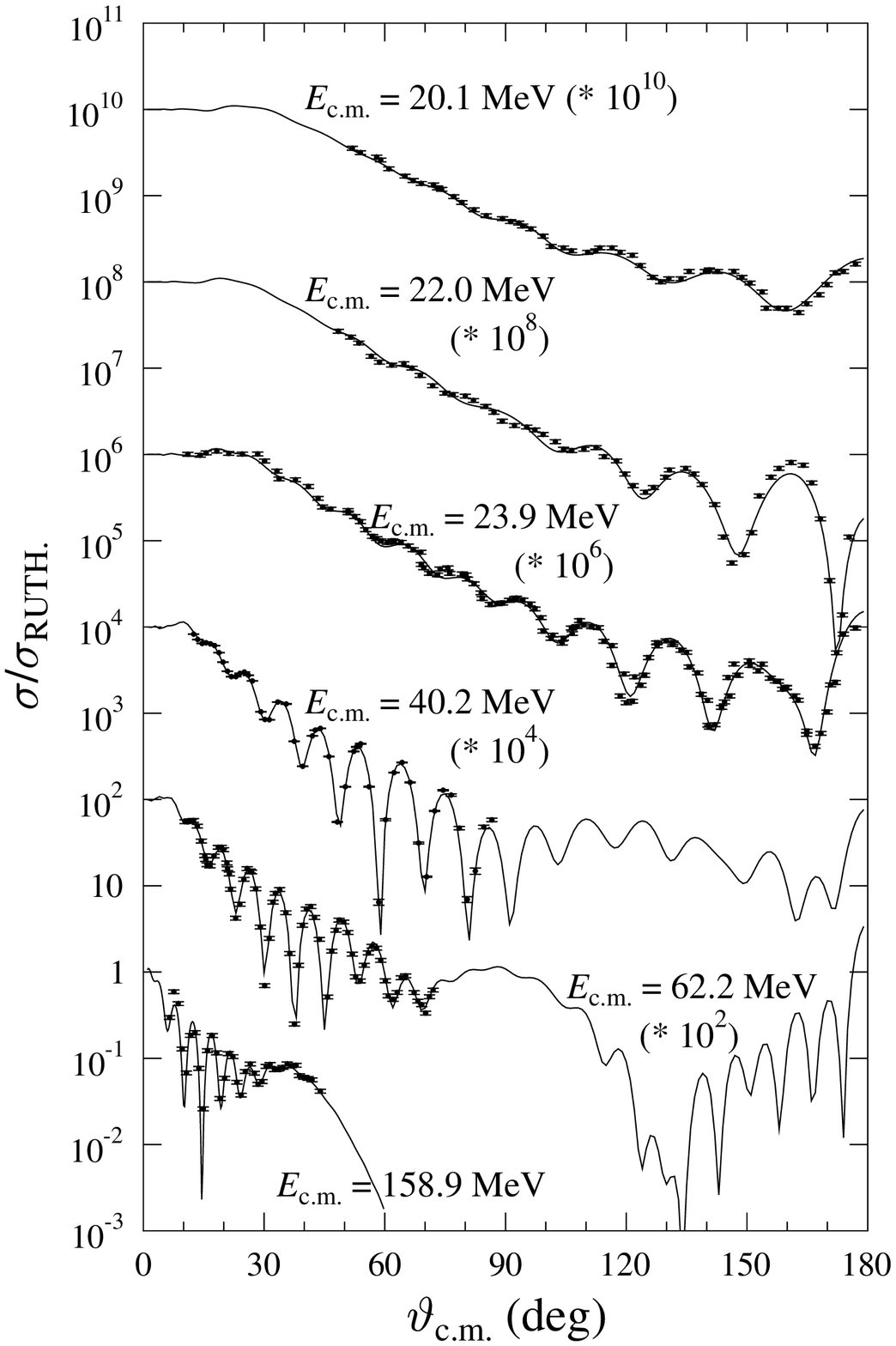}
\caption{
\label{fig:scatall}
Rutherford normalized elastic scattering cross sections of
$^{89}$Y($\alpha,\alpha$)$^{89}$Y reaction at energies between 20 and 160
MeV. The experimental data are taken from literature
\cite{Bri72,Bin69,Als66,eng82,wit75}. The parameters of
the fits are listed in Table \ref{tab:pot}.
}
\end{figure}

The calculated cross sections are compared to the experimental data in the
broad energy range from about 20 MeV up to about 160 MeV in
Fig.~\ref{fig:scatall}. Similar to the result for the new low-energy data
shown in Fig.~\ref{fig:scatlow}, excellent agreement is obtained for all
angular distributions in the wide energy range from slightly above the Coulomb
barrier up to 160 MeV. The parameters of the potential are also listed in
Table \ref{tab:pot}.  

Unfortunately, the angular distributions at $E_{\rm{c.m.}} = 40.2$ MeV and
62.2 MeV do not cover the backward angular range which is most sensitive to
the imaginary part of the potential. On the other hand, the derived potentials
from the analysis of the England {\it et al.}~data at 25 MeV and the corrected
Wit {\it et al.}~data at 25 MeV are in reasonable agreement; this indicates
that the Wit {\it et al.}~data that focus on the backward angular region are
sufficient to derive the optical potential -- provided that the absolute
normalization is correct. In practice this means that data in the forward
region are not necessary to derive the potential but are highly necessary to
define the absolute normalization of the data by comparison to the Rutherford
cross section.

\begin{figure}
\includegraphics[bbllx=0,bblly=45,bburx=540,bbury=680,width=\columnwidth,clip=]{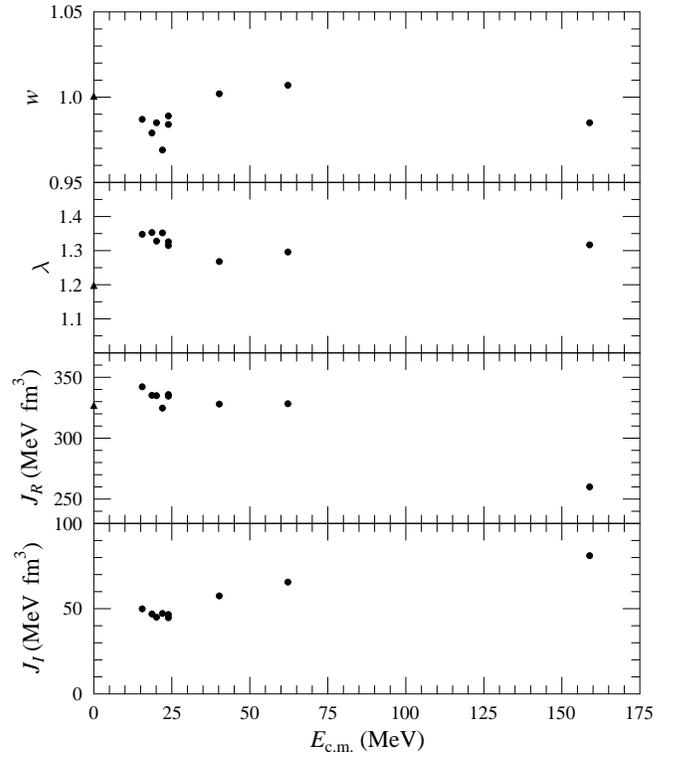}
\caption{
\label{fig:result}
Potential parameters $\lambda$ and $w$ and the integral potential strengths
$J_R$ and $J_I$ for the real and imaginary part of the potentials that are
derived from the fits to the angular distributions in Figs.~\ref{fig:scatlow}
and \ref{fig:scatall} (full circles). Additionally, the result from the bound
state adjustment in Sect.~\ref{sec:cluster} is shown with triangles at $E =
0$. Further discussion see text.
}
\end{figure}

The various available scattering data for $^{89}$Y enable a study of the
energy dependence of the potential parameters that are derived from the fits
to the angular distributions in Figs.~\ref{fig:scatlow} and
\ref{fig:scatall}. For the real part a very smooth dependence of the strength
parameter $\lambda$ and the width parameter $w$ is found, see
Fig.~\ref{fig:result}. As expected, the width parameter $w$ remains close to
unity at all energies, and the strength parameter $\lambda$ is almost
energy-independent and varies less than 10\,\% between 1.27 and 1.35 leading
to volume integrals $J_R$ for the real part in agreement with the systematics
of \cite{atz96}. The much stronger decrease of the real volume integral $J_R$
with energy by about 25\,\% is a consequence of the energy dependence of the
interaction and the dispersion relation which couples the real and imaginary
parts of the potential. (Note that -- as usual -- the negative sign of the
volume integrals $J_R$ and $J_I$ is neglected in the discussion.)

\begin{figure}
\includegraphics[bbllx=15,bblly=30,bburx=440,bbury=375,width=\columnwidth,clip=]{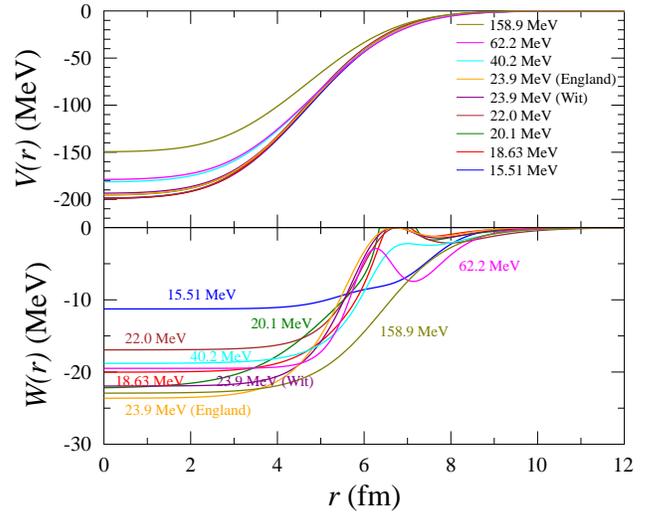}
\caption{
\label{fig:pot}
(Color online) Shape of the potentials for the fits in Figs.~\ref{fig:scatlow}
and \ref{fig:scatall}. The upper part shows the very regular behavior of the
real part $V(r)$. The lower part shows the variations of the shape of the
imaginary potential $W(r)$ at the various energies. Further discussion see
text.
}
\end{figure}

The volume integral $J_I$ of the imaginary part increases with energy, but it remains below $J_I = 100$ MeV\,fm$^3$, in  
agreement with the systematics shown in \cite{atz96}. However, the slope of $J_I$
vs.\ $E_{\rm{c.m.}}$ is not well-defined from the available data, in
particular not in the low-energy region that is most relevant for nuclear
astrophysics. All data between 15 MeV $\le E_{\rm{c.m.}} \le$ 25 MeV lead to
volume integrals of about $J_I \approx 50$ MeV\,fm$^3$. However, a closer look
at the imaginary potentials shows that the shape of the potentials changes
significantly between the different energies (see Fig.~\ref{fig:pot}, lower
part) which complicates the analysis of the $J_I(E)$ dependence. Contrary to
the imaginary potentials, the shape of the real potentials is well-defined
from the folding procedure (see Fig.~\ref{fig:pot}, upper part).

It has to be noted that there are discrete ambiguities for the real part of
the OMP. This has been illustrated e.g.\ in Fig.~5 of \cite{moh97} where 11
so-called ``families'' of potentials were identified which resulted in a
similar description of the low-energy $^{144}$Sm\raa $^{144}$Sm elastic
scattering data. However, these discrete ambiguities are significantly reduced
in the present analysis because of the data at higher energies. It is not
possible to describe the data at 62.2 MeV and 158.9 MeV using potentials from
another family, i.e.\ using a potential strength which is increased or
decreased by about 30\,\%. 

Together with the determination of the bound state potential in
Sect.~\ref{sec:cluster}, the family with J${_R}$ $\approx$ 320 - 350 MeV
fm$^3$ has been selected in the analysis of the low-energy scattering data.
For completeness it has to be pointed out that the above volume integral of
$J_R \approx 320 - 350$ MeV\,fm$^3$ at low energies is only valid for folding
potentials. Slightly different numbers will be obtained if other
parameterizations of the potential, e.g.\ Woods-Saxon potentials, are used
(see e.g.\ \cite{avr03}).

\begin{figure}
\includegraphics[bbllx=25,bblly=35,bburx=440,bbury=240,width=\columnwidth,clip=]{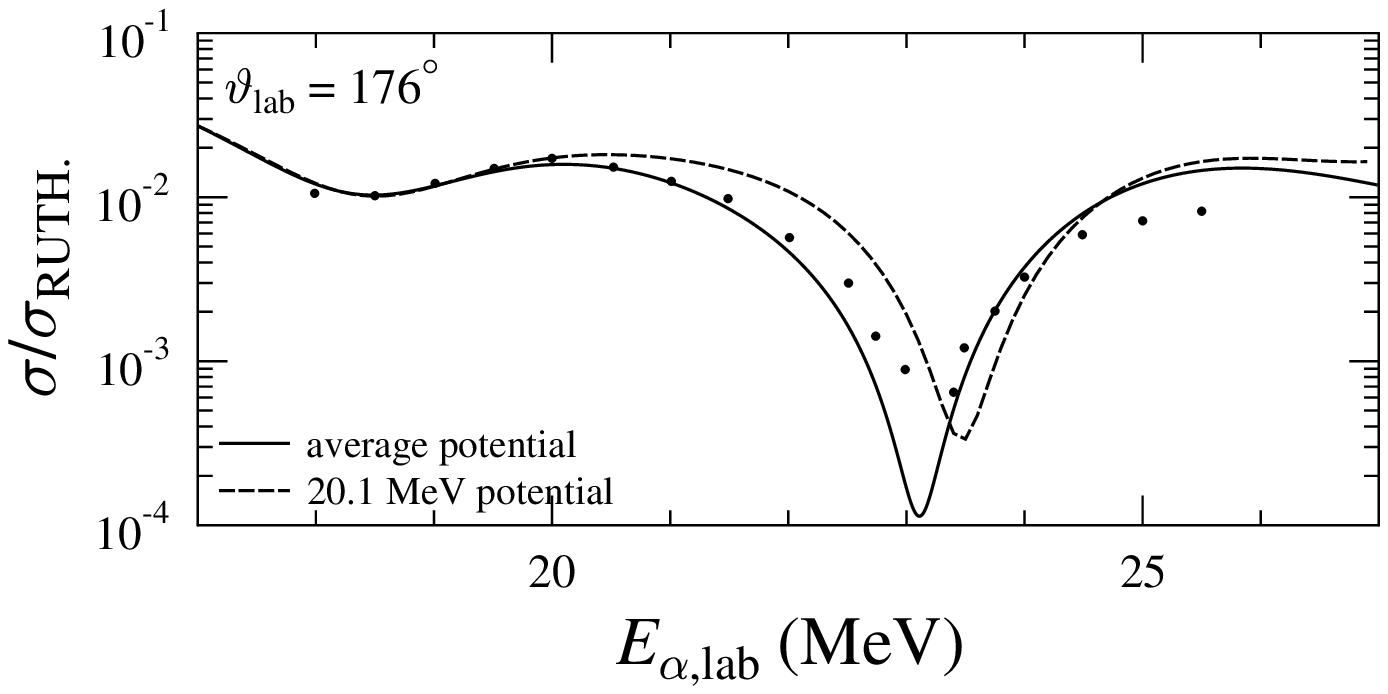}
\caption{
\label{fig:exci}
Excitation function of $^{89}$Y($\alpha$,$\alpha$)$^{89}$Y at the angle
$\vartheta_{\rm{lab}} = 176^\circ$ \cite{wit75} compared to calculations using
the 20.1 MeV potential (dashed line) and an averaged potential from all fits
below 25 MeV (full line). 
}
\end{figure}

In addition to the angular distributions, an excitation function at the very
backward angle of $\vartheta_{\rm{lab}} = 176^\circ$ has been measured in
\cite{wit75} in the energy range of 18 MeV $\le E_{\alpha,{\rm{lab}}} \le$ 26
MeV, i.e.\ covering the energy range of the angular distributions in this work
and in \cite{wit75,eng82}. This excitation function has been calculated using
the potential from the fit to the 20.1 MeV data from \cite{wit75}, see
Fig.~\ref{fig:exci}. Additionally, an averaged potential has been calculated
from all fits below 25 MeV. Both the 20.1 MeV potential and the average
potential are able to reproduce the shape of the excitation function including
the deep minimum around 23 MeV; however, the agreement between theory and
experiment is slightly worse for the excitation function at very backward
angles (compared to the excellent reproduction of the angular
distributions). Because of the relatively weak energy dependence of the volume
integrals $J_R$ and $J_I$ below 25 MeV, this averaged potential will also be
used for comparison with global potentials in Sect.~\ref{sec:global}.

\subsection{$\alpha$-cluster states in $^{93}$Nb = $^{89}$Y $\otimes$
  $\alpha$}
\label{sec:cluster}
$\alpha$ clustering in intermediate and heavy nuclei is a
well-established phenomenon \cite{Buck95,Ohk95,atz96,Ohk98}. A link
between the cluster model and the collective model has been discussed
in \cite{Buck07}, and it has been pointed out that a cluster model
interpretation of collective properties - as recently reviewed in
\cite{Cas07} - emerges.

The
properties of $\alpha$ clustering above shell closures in the
even-even nuclei $^{44}$Ti = $^{40}$Ca $\otimes$ $\alpha$, $^{94}$Mo =
$^{90}$Zr $\otimes$ $\alpha$, and $^{212}$Po = $^{208}$Po $\otimes$
$\alpha$ have been studied extensively
\cite{Buck95,Ohk95,atz96,Ohk98,Buck95b,Ohk98b,Mic00,Buck96,Buck94,Hoy94}.
However, only few studies are available for even-odd nuclei below the
above mentioned shell closures \cite{Mic98,Sak98,Buck92}. Following
the ideas in \cite{Mic98,Sak98} for $^{43}$Sc = $^{39}$K $\otimes$
$\alpha$, we analyze here $\alpha$-cluster properties of $^{93}$Nb =
$^{89}$Y $\otimes$ $\alpha$. The ground state of $^{89}$Y ($Z = 39$,
$N = 50$) has $J^\pi = 1/2^-$ which comes from a one-proton hole in
the $p_{/1/2}$ shell of the neighboring $^{90}$Zr ($Z = 40$, $N =
50$). Counterparts of the well-studied rotational bands in $^{94}$Mo =
$^{90}$Zr $\otimes$ $\alpha$ should be found in $^{93}$Nb = $^{89}$Y
$\otimes$ $\alpha$ with $^{89}$Y = $^{90}$Zr $\otimes$
$p_{1/2}^{-1}$. This study extends a recent review on \al -cluster
states in $N = 50$ $\otimes$ \al\ even-even nuclei \cite{moh08}. 
All excitation energies and decay properties of excited states in $^{93}$Nb
have been taken from \cite{NNDC}.

The full formalism of the applied model can be found in
e.g.\ \cite{atz96,Hoy94}. Here we briefly repeat the important features of the
model. The $\alpha$-cluster wave function can be directly calculated from the
Schroedinger equation and the $\alpha$-nucleus potential which is taken from
the double-folding procedure (see also above). The so-called Wildermuth
condition ensures that the Pauli principle is taken into account:
\begin{equation}
Q = 2N + L = \sum_{i=1}^4 (2n_i + l_i) = \sum_{i=1}^4 q_i
\label{eq:wild}
\end{equation}
where $Q$ is the number of oscillator quanta,
$N$ is the number of nodes and $L$ the relative angular
momentum of the $\alpha$-cluster wave function, and
$q_i = 2n_i + l_i$ are the corresponding quantum numbers
of the nucleons in the $\alpha$ cluster. We take
$q = 4$ and thus $Q = 16$ for the $1/2^-$ band in $^{93}$Nb =
$^{89}$Y($1/2^-$) $\otimes$ $\alpha$. A $1/2^+$
band with $Q = 17$ should also exist in $^{93}$Nb; however, no firm
assignment could be found up to now \cite{NNDC}. Also a higher nodal
band with $Q = 18$ has not yet been found experimentally
\cite{NNDC}. Typical properties of $\alpha$-cluster states are large
reduced widths $\theta_\alpha^2$ or spectroscopic
factors. Unfortunately, no $\alpha$ transfer data like e.g.\ 
($^6$Li,d) on $^{89}$Y can be found in 
\cite{NNDC}; such data have been essential for the assignment of
$\alpha$-cluster properties in many cases \cite{Yam98}.

A prerequisite for a successful description of $\alpha$-cluster states is an
$\alpha$-nucleus potential which is also able to describe the elastic $\alpha$
scattering cross section and/or $\alpha$-decay properties
(e.g.~\cite{Mic98,Buck95,Hoy94}). In the present investigation of $^{93}$Nb we
use a double-folding potential for the real part that is close to the result
of the previous Sect.~\ref{sec:fold}; the imaginary part of the potential
vanishes at the very low energies studied here.

The properties of the $\alpha$-cluster states in $^{93}$Nb with a
$^{89}$Yb($1/2^-$) $\otimes$ $\alpha$ structure have been analyzed in
the following way. In a first step the strength parameter $\lambda$ of
the double-folding potential is adjusted to reproduce the binding
energy of the first $1/2^-$ state in $^{93}$Nb; it is located at $E_x
= 30.8$\,keV or $E = Q_\alpha - E_x = -1900.7$\,keV with the $\alpha$
binding energy $Q_\alpha = -1931.5$\,keV. The result $\lambda =
1.1951$ and the corresponding volume integral $J_R =
326.1$\,MeV\,fm$^3$ are very close to the numbers obtained from the
analysis of the scattering data. For simplicity, the width parameter $w$ has
been fixed at $w = 1$ for the bound state calculations.

For a perfect rotator the calculations with a realistic potential should be
able to predict the energies of all members of this rotational band. However,
it has been shown that the strength of folding potentials has to be slightly
reduced with increasing angular momentum (or increasing excitation energy).
It was found that an excellent description of the energies within rotational
bands is obtained using a potential strength with a weak dependence on the
angular momentum $L$:
\begin{equation}
\lambda(L) = \lambda(L = 0) - c \times L
\label{eq:laml}
\end{equation}
with small values for the constant $c \approx (3-5) \times 10^{-3}$
\cite{Mic98,Ohk98b}. Thus, in a second step the potential strength was
adjusted to the centroid of the $3/2^-$ and $5/2^-$ states with $L =
2$; we find $\lambda = 1.1856$ and derive $c = 4.75 \times 10^{-3}$ in
excellent agreement with several other nuclei \cite{Mic98}. Then we
use a weak spin-orbit potential $V_{LS} \sim 1/r \times dV_F/dr$; the
strength of this spin-orbit potential is adjusted to reproduce the
splitting of the $3/2^-$ and $5/2^-$ states with $L = 2$.
Now all parameters of the potential are fixed, and it is possible to
calculate the energies of the members of the $Q = 16$ rotational band
in $^{93}$Nb. The results are listed in Table \ref{tab:cluster}.
\begin{table}
  \caption{\label{tab:cluster} 
  Properties of the $Q = 16$, $K^\pi = 1/2^-$ rotational band in
  $^{93}$Nb. For $L \ge 6$ only the centroid of the two states with $J
  = L \pm 1/2$ is given. Additionally the excitation energy of the $1/2^+$,
  $3/2^+$ band head of the $Q = 17$, $K^\pi = 1/2^+$ band is
  predicted. 
}
\begin{center}
\begin{tabular}{cccrrr}
$L$
& $J^\pi$
& $Q$
& \multicolumn{1}{c}{$E$}
& \multicolumn{1}{c}{$E_x(^{93}{\rm{Nb}})$}
& \multicolumn{1}{c}{$\lambda$} \\
$-$
& $-$
& $-$
& \multicolumn{1}{c}{(keV)}
& \multicolumn{1}{c}{(keV)}
& \multicolumn{1}{c}{$-$} \\
\hline
0  & $1/2^-$            & 16 & $-1901$   
& 31\footnotemark[1]       & 1.1951\footnotemark[1] \\
2  & $3/2^-$            & 16 & $-1244$   
   & 687\footnotemark[1]   & 1.1856\footnotemark[1] \\
2  & $5/2^-$            & 16 & $-1121$   
   & 810\footnotemark[1]    & 1.1856\footnotemark[1] \\
4  & $7/2^-$            & 16 & $-457$   
   & 1474\footnotemark[2]    & 1.1761\footnotemark[2] \\
4  & $9/2^-$            & 16 & $-238$   
   & 1693\footnotemark[2]    & 1.1761\footnotemark[2] \\
6  & $11/2^-$, $13/2^-$ & 16 & $+613$    
   & 2545\footnotemark[2]    & 1.1666\footnotemark[2] \\
8  & $15/2^-$, $17/2^-$ & 16 & $+1681$   
   & 3613\footnotemark[2]    & 1.1571\footnotemark[2] \\
10 & $19/2^-$, $21/2^-$ & 16 & $+2878$   
   & 4809\footnotemark[2]    & 1.1476\footnotemark[2] \\
12 & $23/2^-$, $25/2^-$ & 16 & $+4221$   
   & 6152\footnotemark[2]    & 1.1381\footnotemark[2] \\
14 & $27/2^-$, $29/2^-$ & 16 & $+5749$   
   & 7680\footnotemark[2]    & 1.1286\footnotemark[2] \\
16 & $31/2^-$, $33/2^-$ & 16 & $+7514$   
   & 9445\footnotemark[2]    & 1.1191\footnotemark[2] \\
1  & $1/2^+$, $3/2^+$   & 17 & $+4765$   
   & 6696\footnotemark[2]    & 1.1904\footnotemark[2] \\
0  & $1/2^-$            & 18 & $+9432$   
& 11363\footnotemark[2]       & 1.1951\footnotemark[2] \\
\hline
\end{tabular}
\footnotetext[1]{
  $E$ from experimental data; $\lambda$ adjusted to fit the energy $E$.
}
\footnotetext[2]{
  $E$ predicted using $\lambda(L) = \lambda(L=0) - c \times L$,
  Eq.~(\ref{eq:laml}). 
}
\end{center}
\end{table}

There are a number of candidates for the $L = 4$ states with $J^\pi =
7/2^-$ and $9/2^-$ at $E_x = 1364$\,keV, 1500\,keV, 1603\,keV, and
1916\,keV; however, no firm assignment is possible at the present
stage. Therefore, for the higher-lying members of the rotational band
with $L \ge 6$
only the centroid energy of the states with $J = L \pm 1/2$ are listed
in Table \ref{tab:cluster}. In addition, the band heads of the $L =$
odd, $Q = 17$ and $L =$ even, $Q = 18$ bands are predicted.

The uncertainties of the predictions in Table \ref{tab:cluster} can be
estimated from the uncertainty of the potential strength parameter
$\lambda(L)$ and thus from the uncertainty of the constant $c$ of
about 20\,\% \cite{Mic98}. The uncertainty of the potential strength
$\lambda(L)$ in Eq.~(\ref{eq:laml}) increases with $L$ from about $2
\times 10^{-3}$ at $L = 4$ to about $0.01$ around $L = 10$. The
resulting uncertainty is of the order 
of about 150\,keV for the $L = 4$ states and up to several hundred keV
for states with higher angular momentum $L$. Together with an
additional uncertainty from the weak spin-orbit potential of the order
of 100\,keV the total uncertainty for the predicted excitation
energies is about 200\,keV for the $L = 4$ states and larger for
higher $L$. Slightly varying potentials have been used for the
description of bands with different $Q$. This leads to an increased
uncertainty  of the order of 1\,MeV for the band heads of the $Q = 17$
and $Q = 18$ rotational bands in $^{93}$Nb.

Electromagnetic decay properties may provide additional hints for the
$\alpha$-cluster structure of the $Q = 16$ rotational band in
$^{93}$Nb. It is straightforward to calculate the wave functions of
these states and the reduced transition strengths $B(E2,J^\pi_i
\rightarrow J^\pi_f)$ for transitions from an initial state with
$J^\pi_i$ to a final state with $J^\pi_f$ \cite{Buck77}. The results
are listed in Table \ref{tab:be2}. Indeed, enhanced transition
strengths of the order of 10\,W.u.\ are found for several transitions.
\begin{table*}
  \caption{\label{tab:be2} 
  Calculated reduced transition strengths $B(E2)$ in the $Q = 16$, $K^\pi =
  1/2^-$ rotational band in $^{93}$Nb compared to experimental values
  \cite{NNDC}.
}
\begin{center}
\begin{tabular}{ccrccrrr}
$L_i$
& $J^\pi_i$
& \multicolumn{1}{c}{$E_{x,i}$}
& $L_f$
& $J^\pi_f$
& \multicolumn{1}{c}{$E_{x,f}$}
& \multicolumn{1}{c}{$B(E2,J_i\rightarrow J_f)_{\rm{calc}}$}
& \multicolumn{1}{c}{$B(E2,J_i\rightarrow J_f)_{\rm{exp}}$} \\
$-$
& $-$
& \multicolumn{1}{c}{(keV)}
& $-$
& $-$
& \multicolumn{1}{c}{(keV)}
& \multicolumn{1}{c}{(W.u.)}
& \multicolumn{1}{c}{(W.u.)} \\
\hline
2  & $3/2^-$  &  687   & 0   & $1/2^-$  &   31  & 8.6  & $11^{+35}_{-10}$ \\
2  & $5/2^-$  &  810   & 0   & $1/2^-$  &   31  & 8.6  & $< 79$ \\
4  & $7/2^-$  & 1474   & 2   & $3/2^-$  &  687  & 10.7 & $-$ \\
4  & $7/2^-$  & 1474   & 2   & $5/2^-$  &  810  & 1.2  & $-$ \\
4  & $9/2^-$  & 1693   & 2   & $5/2^-$  &  810  & 12.0 & $-$ \\
\hline
\end{tabular}
\end{center}
\end{table*}

Experimental data for transition strengths are rare and have huge
uncertainties. Within these uncertainties, the transition strengths
for the transitions from the $L = 2$ states with $J^\pi_i = 3/2^-$ and
$5/2^-$ to the final state with $L = 0$, $J^\pi_f = 1/2^-$ are
calculated correctly.

For transitions from $L = 4$ states to $L = 2$ states predictions are
given in Table \ref{tab:be2}. It is interesting to note that the
absolute transition strengths $\Gamma_{i \rightarrow f}$ depend
sensitively on the predicted energies of these states because of the
$E_\gamma^5$ dependence of $E2$ transitions. However, the reduced
transition strengths are almost independent of the predicted
energy. From the above mentioned candidates for the $L = 4$ states,
only the transition from the $J_i = 7/2$ state at $E_x = 1500$\,keV to
the $J^\pi_f = 5/2^-$ state at $E_x = 810$\,keV has been detected
experimentally. The preferred electromagnetic decay mode of most of
the $L = 4$ candidates proceeds via $E1$ transitions to low-lying
$J^\pi_f = 7/2^+$ and $9/2^+$ states which can be found at $E_x =
0$\,keV, 744\,keV, and 1083\,keV. Thus, the predictions of $B(E2)$
values in Table \ref{tab:be2} are not able to firmly assign the $L =
4$ members of the $Q = 16$ rotational band because of missing
experimental data for weak decay branches. It has to be noted that a
1\,MeV transition in $^{93}$Nb with a $E1$ strength of
$10^{-3}$\,W.u.\ corresponds to $\Gamma_\gamma(E1) \approx 1.4$\,meV
whereas a similar 1\,MeV $E2$ transition with 10\,W.u.\ corresponds
to a much smaller radiation width of only $\Gamma_\gamma(E2) =
0.2$\,meV. 

In a recent paper, the state at $E_x = 1500$\,keV has been assigned $J^\pi =
9/2^-$ \cite{Orce07} instead of $J^\pi = 7/2^{(-)}$ \cite{NDS,Avc82}, and an
E2 transition strength of $26.4^{+9.7}_{-6.2}$\,W.u.\ has been measured for
the transition to the $5/2^-$ state at $E_x = 810$\,keV \cite{Orce07}. A tiny
readjustment of the potential strength parameter $\lambda$ of about 0.2\,\%
from the extrapolated value of 1.1761 to 1.1788 is required to shift the
$9/2^-$ state from the predicted value $E_x = 1693$\,keV to
$1500$\,keV. Simultaneously, the $7/2^-$ state is shifted from the prediction
$E_x = 1474$\,keV to $1282$\,keV which is very close to the $J^\pi =
5/2^-,7/2^-$ state at $E_x = 1364$\,keV. As pointed out above, the calculated
transition strength of 12\,W.u.\ for the $9/2^- \rightarrow 5/2^-$ E2
transition is practically independent of the precise excitation energies. The
enhanced experimental transition strength confirms the $J^\pi = 9/2^-$
assignment of \cite{Orce07} for the $E_x = 1500$\,keV state and the
interpretation as a member of the $L = 4$ doublet with $^{89}$Y $\otimes$
\al\ structure.

\section{Global Optical Model predictions}
\label{sec:global}
After the successful study of $^{89}$Y(\al ,\al )$^{89}$Y elastic scattering
and $^{93}$Nb = $^{89}$Y $\otimes$ \al\ bound state properties with a locally
optimized potential we will now test the predictions of global \al -nucleus
optical potentials for the $^{89}$Y(\al ,\al )$^{89}$Y elastic scattering
cross section. Obviously, the best description of elastic scattering data is
obtained from a locally adjusted potential. However, such a locally optimized
potential requires experimental data for adjustment which are often not
available. In particular, for unstable nuclei it is not possible to measure
elastic scattering data with the required precision. Therefore global
$\alpha$-nucleus potentials are required for the prediction of scattering and
reaction data. In the following we will compare the predictions of published
global potentials to our experimental results.

The prediction of cross sections for unstable nuclei requires the
extrapolation of global potentials into regions of the nuclear chart where no
scattering data are available to test the extrapolations. It has been shown
recently \cite{gal05} that the variation of the potential along an isotopic
chain (in that case $Z = 50$) can be sensitively studied by a comparison of
the scattering cross sections of a neutron-rich and a neutron-deficient
isotope. Here we extend this idea and study the variation of the scattering
cross sections along the $N = 50$ isotonic chain by a comparison of scattering
data for $^{89}$Y and $^{92}$Mo.

Several different parameterizations for the optical potential
exist. For reactions involving alpha particles the optical potential is
described mainly by complex Woods-Saxon potentials. Some authors investigated the use of
higher order terms of the Woods-Saxon function, too
\cite{mic77, gub81}. Model-independent parameterizations have also been
studied, either with spline functions \cite{mic77, put77} or series of
Fourier-Bessel functions added to the Woods-Saxon parameterizations
\cite{fri78, gil80} or with a sum of Gaussians \cite{gub81}.  A folding model
has been introduced by Kobos \textsl{et al.} \cite{kob84}. In the framework of
this model an energy- and density-dependent effective interaction --- approximating the nonlocal exchange component by an empirical parameterization --- was used to describe alpha-particle elastic scattering at energies from 25 to 120
MeV. This model was also extended for inelastic scattering. However, in all of these investigations the parameterization of the
imaginary part of the optical potential varied from nucleus to nucleus, in
order to obtain a good description of the scattering cross sections for all
the nuclei studied.

In the framework of the $p$ process network calculations a large number of
reactions involving alpha particles (alpha-induced reactions and alpha
particle emission) has to be taken into account. As the $p$ process path is
located in a region of unstable nuclei on the neutron-deficient side of the
chart of nuclides, experimental data are practically not available to adjust
potential parameters of the $\alpha$-nucleus potential.  Therefore, a global
$\alpha$-nucleus optical potential is required for the theoretical prediction
of reaction cross sections involving $\alpha$ particles within the statistical
Hauser-Feshbach model. 

Several different global and regional parameterizations have been developed in
recent years to describe the interaction between nuclei and alpha
particles. In the following an overview of the parameterizations is given
which are studied in this work. 

($i$) The regional $\alpha$-nucleus potential of Avrigeanu \textsl{et
  al.}~\cite{avr07} corresponds to average mass--, charge--, and
energy--dependent Woods-Saxon parameters based on local potentials obtained by
analysis of 108 experimental angular distributions of $\alpha$-particle
elastic--scattering on target nuclei from $^{50}$Ti to $^{124}$Sn and
$\alpha$-particle energies from 8.1 to 49\,MeV.  The local Woods-Saxon
parameter sets provided by the analysis of the new experimental data (Table 2
of Ref. \cite{avr07}) have led to similar results as those shown in
Fig.~\ref{fig:scatlow} (see also Fig.~7 of Ref.~\cite{avr07}).

($ii$)
The recently published global potential by Kumar \textsl{et al.}~\cite{kum06}
claims to describe data in a wide mass ($12 \le A \le 209$) and energy region
(from the Coulomb barrier up to 140\,MeV); however, problems below 30\,MeV had
to be compensated by an enhanced imaginary volume integral, and even with this
enhancement the description of $^{90}$Zr(\al ,\al )$^{90}$Zr scattering at
15\,MeV is reasonable but not excellent (see Fig.~2 in \cite{kum06}). A
similar reproduction of the $^{89}$Y(\al ,\al )$^{89}$Y scattering data below
20\,MeV is expected. 

($iii$) The widely used potential by McFadden and Satchler
\cite{mcf} is a very simple 4-parameter Woods-Saxon potential with mass- and
energy-independent parameters. Despite its simplicity it provides an excellent
description of \al -scattering data and cross sections of \al -induced
reactions; e.g., this potential is used as default for the Hauser-Feshbach
calculations of astrophysical reaction rates by Rauscher and Thielemann
\cite{rau01,NONSMOKER}. 

Unfortunately, the latest version \cite{dem09,dem07} of the
potentials by Demetriou {\it et al.}~\cite{dem02} are only published in
conference proceedings and cannot be used without the authors of
\cite{dem09,dem07}; we do not intend to show results from the early and
perhaps out-dated potentials in \cite{dem02}.

\begin{figure}
\resizebox{0.8\columnwidth}{!}{\rotatebox{270}{\includegraphics[clip=]{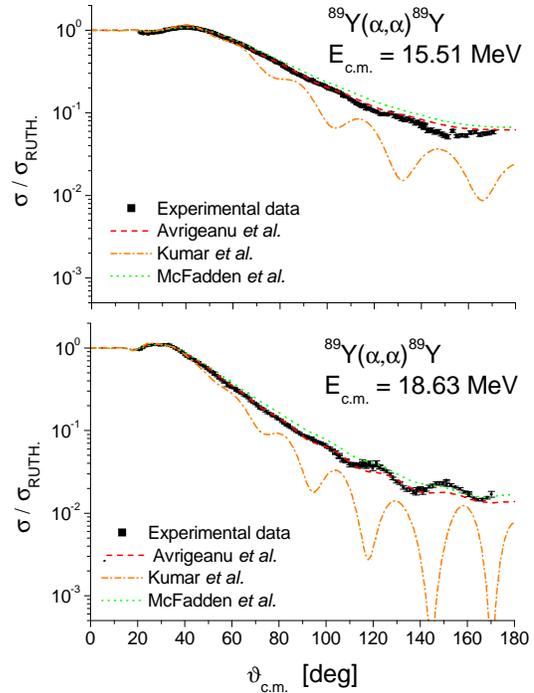}}}
\caption{
\label{fig:ruth}
(color online). Rutherford normalized elastic scattering cross sections of $^{89}$Y($\alpha,\alpha$)$^{89}$Y reaction at E$_{c.m.}$ = 15.51 and 18.63 MeV versus the angle in center-of-mass frame. The lines correspond to the predictions using different global optical potential parameter sets. For more information see Sec.~\ref{sec:global}.} 
\end{figure}

\subsection{Angular distributions: comparison with theoretical models}
\label{sec:angular}
\begin{figure}
\resizebox{0.8\columnwidth}{!}{\rotatebox{270}{\includegraphics[clip=]{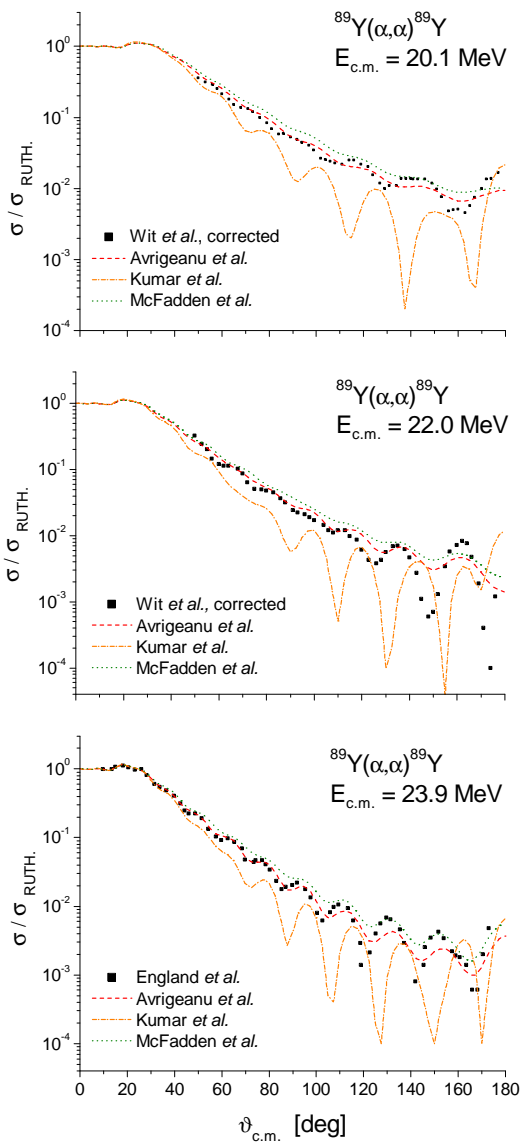}}}
\caption{\label{fig:wit_ang}
(color online). Rutherford normalized elastic scattering cross sections of
$^{89}$Y($\alpha,\alpha$)$^{89}$Y reaction at E$_{c.m.}$ = 20.1, 22.4 and 23.9
MeV versus the angle in center-of-mass frame taken from Wit \textit{et al.}
and England \textit{et al.}~\cite{wit75, eng82}. The data by Wit \textit{et
  al.}~\cite{wit75} have been corrected by a factor of 1.45 to come into
agreement with the England \textit{et al.}\ data \cite{eng82} at 23.9 MeV. The
lines correspond to the predictions using different global optical
potential parameter sets. For more information see Sec.~\ref{sec:global}.}
\end{figure}

The regional optical potential (ROP) of Avrigeanu \textsl{et al.}
\cite{avr03} was obtained by a semi-microscopic analysis, using the double
folding model \cite{kho94}, based on alpha-particle elastic scattering on A
$\approx$ 100 nuclei at energies below 32 MeV. The energy-dependent
phenomenological imaginary part of this semi-microscopic optical potential
takes into account also a dispersive correction to the microscopic real
potential. A small revision of this ROP and especially the use of local
parameter sets were able to describe the variation of the elastic scattering
cross sections along the Sn isotopic chain \cite{avr_ad}. A further step to
include all available $\alpha$-induced reaction cross sections below the
Coulomb barrier has recently been carried out \cite{avr07}. First, the ROP
based entirely on $\alpha$ particle elastic scattering \cite{avr03} was
extended to $A\sim$ 50-120 nuclei and energies from $\sim$ 13 to 50
MeV. Secondly, an assessment of available $(\alpha,\gamma)$, $(\alpha,n)$ and
$(\alpha,p)$ reaction cross sections on target nuclei ranging from $^{45}$Sc
to $^{118}$Sn at incident energies below 12 MeV was carried out. In this work
the most recent potential of Avrigeanu \textsl{et al.} \cite{avr07}, is used
to calculate the scattering cross sections.

\begin{table}
\caption{
$\chi^2_{red}$ of predictions using different global parameterizations
  compared with the angular distributions derived in the present work. No
  parameters have been adjusted to the new experimental data.
}
\setlength{\extrarowheight}{0.1cm}
\begin{ruledtabular}
\begin{tabular}{ccccccc}
\parbox[t]{4.0cm}{\centering{Global Parameterizations }} &
\multicolumn{2}{c}{$^{89}$Y($\alpha,\alpha$)$^{89}$Y}  
&
\parbox[t]{1.6cm}{\centering{Ref.}} \\
\hline
\parbox[t]{1.6cm}{\centering{   }} &
\parbox[t]{1.6cm}{\centering{ 15.51 MeV}} &
\parbox[t]{1.6cm}{\centering{ 18.63 MeV}} \\
Avrigeanu & 6.5  & 5.5  &    \cite{avr07} \\
Kumar     & 181  & 287  &    \cite{kum06} \\
McFaddden & 35.3 & 40.9 &    \cite{mcf} \\
\end{tabular}
\end{ruledtabular}
\end{table}

The potential from Kumar \textsl{et al.}~\cite{kum06} was proposed to describe
alpha-induced reactions on 12 $\leq$ A $\leq$ 209 target nuclei at E $\leq$
140 MeV. In that work, the systematics of volume integrals has been used to
determine the real and the imaginary parts of the potential. The real
potential volume integrals have been taken from the work of Atzrott \textsl{et
  al.} \cite{atz96}.  The best fit volume integrals from the phenomenological
analysis, consistent with the above set, had been employed in their
analysis. A similar approach was followed for the imaginary part starting with
the volume integral systematics of the imaginary part. Moreover at energies
below 30 MeV a dispersive correction is taken into account. Calculations using
this potential are able to reproduce the measured alpha scattering angular
distributions at higher energies \cite{kum06}. Here we compare the predicted
angular distributions from the Kumar \textsl{et al.} \cite{kum06} potential to
our new experimental data at lower energies close to the Coulomb barrier.

For completeness, we take into account also the potential of McFadden
\textsl{et al.} \cite{mcf}. Numerous alpha elastic scattering experiments were
done on nuclei between O and U at alpha energies of 24.7 MeV in the 60`s. Fits
were obtained using a four-parameter Woods-Saxon potential.

In Figure~\ref{fig:ruth} the measured angular distributions of the
$^{89}$Y($\alpha,\alpha$)$^{89}$Y elastic alpha scattering at E$_{c.m.}$=
15.51 and E$_{c.m.}$=18.63 MeV are shown. The different lines correspond to
the predictions using the above discussed global and regional optical
potential parameterizations without any further adjustment of parameters. The
overall agreement between the calculations performed with using the potentials
of Avrigeanu and McFadden and the experimental data is good. Only the
calculation performed with the potential of Kumar \textsl{et al.} gives a
significantly worse description of the experimental data. For a strict
comparison between the potentials of Avrigeanu, and McFadden a $\chi^2$
analysis has been done. The resulting $\chi^2$ parameters can be found in
Table I. As can be seen, the quality of the different parameterizations is
similar although a value of $\chi^2_{red} \approx 1$ cannot be reached by any
of the global potentials.

Our analysis is extended up to 23.9 MeV using the data of Wit \textsl{et al.}
\cite{wit75} and England \textsl{et al.} \cite{eng82}. The
$^{89}$Y($\alpha,\alpha$)$^{89}$Y elastic scattering was studied by Wit
\textsl{et al.} at E$_{c.m.}$ $\approx$ 20.1, 22.4, and 23.9 MeV and by
England \textsl{et al.} at E$_{c.m.}$ $\approx$ 23.9 MeV. Note that we have
applied the same reduction of a factor of 1.45 to all experimental data on
$^{89}$Y in the Wit \textsl{et al.} paper \cite{wit75} (see
Sect.~\ref{sec:local}).

In general, both the Avrigeanu and McFadden parameterizations can describe roughly both the
magnitude and the oscillation pattern of the angular distributions measured at
energies between 20.1 and 23.9 MeV. The calculations performed with the Kumar potential overestimate the
strength of the oscillation, predicting deeper minima at backward angles. In
the case of calculations performed with the other two global
parameterizations slight differences at backward angles between the measured
and calculated data can be seen in Fig. \ref{fig:wit_ang}.

\subsection{Excitation functions}
\begin{figure}
\resizebox{0.9\columnwidth}{!}{\rotatebox{0}{\includegraphics[clip=]{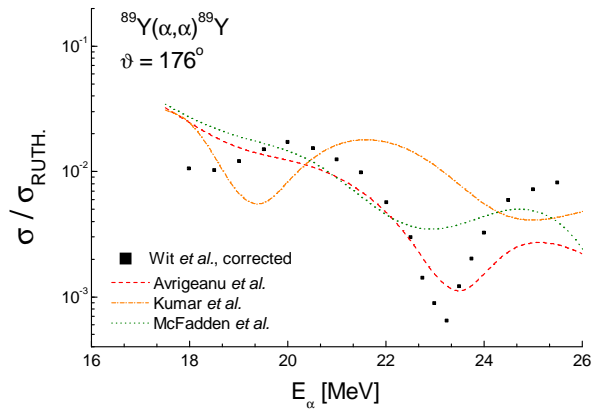}}}
\caption{\label{fig:wit_exc} (color online). Excitation functions for elastic $\alpha$ scattering on $^{89}$Y 
target nuclei at a scattering angle of 176$^{\circ}$ taken from the 
literature \cite{wit75} and corrected by a factor of 1.45 (see text in Sect.~\ref{sec:angular}). The different curves correspond to the 
predictions using different global and regional potentials. For more information see Sec.~\ref{sec:global}.}
\end{figure}

A perfect global optical potential should be able to predict angular
distributions including the diffraction-like patterns which show up at
backward angles with increasing energy. It has to be pointed out that the
backward angular region has a higher sensitivity to the optical potential than
the forward angular region where the Coulomb interaction is dominating. For a
particular study of this backward angular region, an excitation function has
been measured by Wit \textsl{et al.} \cite{wit75} at $\vartheta =
176^\circ$.

The measured excitation function of alpha particle elastic scattering on
$^{89}$Y at a scattering angle 176$^\circ$ \cite{wit75} and the corresponding
optical model calculations using the potential from Avrigeanu \textsl{et al.}
\cite{avr07}, Kumar \emph{et al.} \cite{kum06} and McFadden and Satchler
\cite{mcf} are shown in Fig.~\ref{fig:wit_exc}. Although there is reasonable
agreement with the magnitude of the cross section in all calculations, the
global potentials are not able to reproduce the energy dependence of the cross
section. The sharp minimum of the experimental excitation function around 23
MeV is rather well described by the potential of Avrigeanu \textsl{et al.}
\cite{avr07}, suggested by the global potential of McFadden and Satchler, and
not predicted by the Kumar \textsl{et al.} global potential. The appropriate
description of particular features of the experimental elastic scattering of
alpha particles on $^{89}$Y proves conclusively the suitable account of the
nuclear absorption merely by the potential of Avrigeanu \textsl{et al.}
However, the small scattering cross section at very backward angles (in
particular in the pronounced minimum around 23 MeV in the excitation function,
see Fig.~\ref{fig:exci}) may be affected by small compound
contributions \cite{Cze08}. Thus, minor deviations between the experimental
and calculated cross sections are acceptable.

\subsection{Comparison of global and local potentials}
\label{sec:comp}
For further improvement of the global \al -nucleus potentials a deeper
understanding of the differences between the potentials and the resulting
cross sections is required. In the following paragraphs we will compare the
shapes of the different potentials and the elastic phase shifts $\delta_L$ and
scattered wave amplitudes $\eta_L$ that define the elastic scattering cross
section. Note that the scattering cross section is related to the sum of the
absolute squares over all contributing partial waves; the study of the
underlying $\delta_L$ and $\eta_L$ will allow a better understanding of the
global potentials. However, it is beyond the scope of the present paper to
derive a new and improved global \al -nucleus potential.

\begin{figure}
\includegraphics[bbllx=15,bblly=30,bburx=440,bbury=375,width=\columnwidth,clip=]{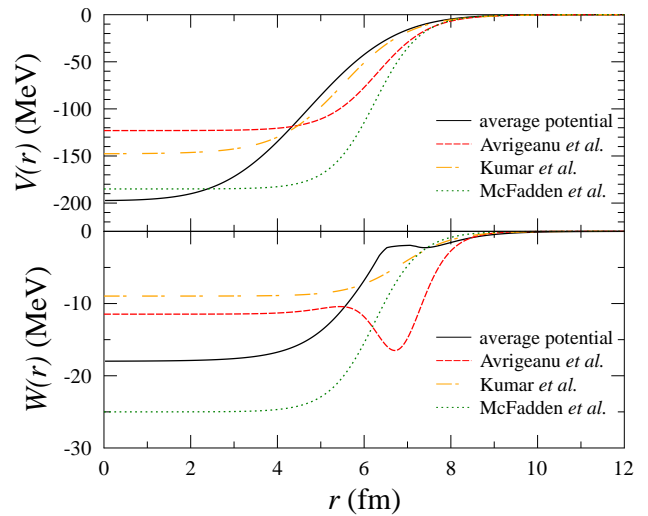}
\caption{
\label{fig:potglob}
(Color online) Comparison between the averaged local potential (see
Sect.~\ref{sec:fold}) and the global parameterizations at 18.63 MeV (see Sect.\ref{sec:global}).
}
\end{figure}

The shapes of the real and imaginary part of the local potential have already
been shown in Fig.~\ref{fig:pot}. Because of the minor energy dependence of
the volume integrals $J_R$ and $J_I$ below 30\,MeV (see
Fig.~\ref{fig:result}), an averaged local potential is used for comparison to
the global potentials. The global potentials are shown in
Fig.~\ref{fig:potglob}. Huge differences can be seen for small radii,
i.e.\ the nuclear interior, whereas at larger radii ($r > 8$\,fm) all
potentials are close to each other. %\textsl{The integral parameters of the potentials in Fig.~\ref{fig:potglob} are listed in Table \ref{tab:newtable}.}

\begin{figure}
\includegraphics[bbllx=15,bblly=30,bburx=440,bbury=375,width=\columnwidth,clip=]{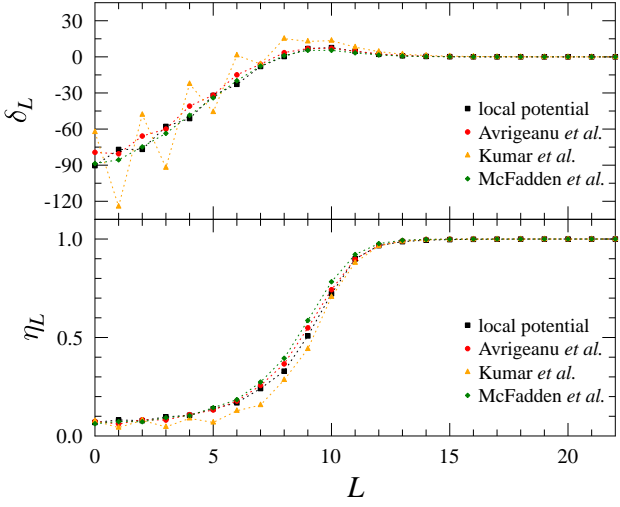}
\caption{
\label{fig:etaphase}
(Color online) Comparison between the local potential (see
Sect.~\ref{sec:fold}) and the global parameterizations (see
Sect.\ref{sec:global}) for $E = 18.63$\,MeV: 
elastic scattering phase shift $\delta_L$ (upper) and amplitude of the
scattered wave $\eta_L$ (lower) in dependence of the angular momentum $L$. The
data points are connected by dashed lines to guide the eye. Note that the
scale in $L$ with $0 \le L \le 22$ corresponds roughly the scale in $r$ of
Fig.~\ref{fig:potglob} with $0 \le r \le 12$\,fm according to Eq.~(\ref{eq:L}).
}
\end{figure}

In a semi-classical picture larger radii in the potential correspond to larger
impact parameters that are related to the angular momentum $\vec{L} = \vec{r}
\times \vec{p}$. E.g., for the energy $E = 18.63$\,MeV this leads to the
approximate relation
\begin{equation}
|\vec{L}|/\hbar \approx 1.85 \times r/{\rm{fm}}
\label{eq:L}
\end{equation}
As expected, the scattering phase shifts $\delta_L$ and scattered wave
amplitudes $\eta_L$ are in agreement for $L \gtrapprox 15$ (see
Fig.~\ref{fig:etaphase}). However, the nuclear potential is relatively weak at
such large radii, and thus the phase shifts $\delta_L$ are small and $\eta_L
\approx 1$ (no absorption); the scattering cross section is dominated by the
Coulomb interaction.

The situation changes for smaller angular momenta. Around $L \approx 10$ (or
semi-classically, $r \approx 5.4$\,fm) the decreasing $\eta_L$ indicate
increasing absorption. Simultaneously, the $\delta_L$ start to deviate from
zero. For small $L \le 5$ (corresponding $r \le 2.7$\,fm) the $\eta_L$ values
are coming close to zero, i.e.\ full absorption of the respective partial
wave. For these partial waves the calculated phase shift $\delta_L$ does not
have strong impact on the scattering cross section. In other words, the
elastic scattering cross section is not very sensitive to the potential at
small radii. However, one should keep in mind that this simple semi-classical
interpretation is not strictly valid; the determination of $\delta_L$ and
$\eta_L$ from the solution of the Schr\"odinger equation depends on the
underlying real and imaginary potentials $V(r)$ and $W(r)$ for the whole range
in $r$ where $V(r)$ or $W(r)$ deviate from zero.

From Fig.~\ref{fig:etaphase} it can be seen that the potential by Avrigeanu
{\it et al.} provides $\delta_L$ and $\eta_L$ that are close to the result of
the local analysis (which is taken as a reference here) for all $L$. In
particular, the most relevant partial waves around $L \approx 10$ are nicely
reproduced. The McFadden/Satchler potential leads to slightly but
systematically larger values of $\eta_L$ for the most relevant $L$ around $L
\approx 10$. This weaker absorption of the McFadden/Satchler potential can
also be seen in Figs.~\ref{fig:ruth} and \ref{fig:wit_ang} where the
calculated cross section in the backward region overestimates the experimental
data. The phase shifts for small $L$ are best reproduced by McFadden/Satchler;
however, as pointed out above, the scattering cross section is not very
sensitive to this region. (This may be different for the calculation of \al
-induced reaction cross sections where the McFadden/Satchler potential has
been used very successfully!) The predictions of $\delta_L$ and $\eta_L$
from the Kumar {\it et al.}  potential deviate significantly from the results
of the local potential, and consequently the calculated scattering cross
sections are not in good agreement with the experimental data.

\subsection{Variation of the scattering cross section along N = 50 isotonic chain}
\label{sec:isotonic}

\begin{figure}
\resizebox{0.7\columnwidth}{!}{\rotatebox{270}{\includegraphics[clip=]{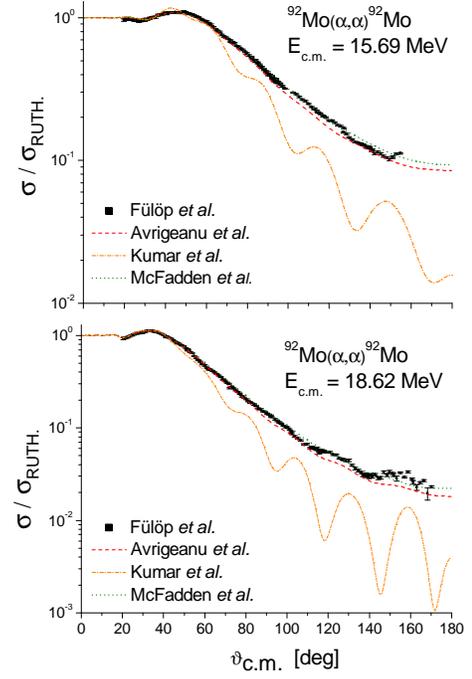}}}
\caption{
\label{fig:mo} (color online). 
Rutherford normalized elastic scattering cross sections of $^{92}$Mo($\alpha,\alpha$)$^{92}$Mo reaction at E$_{c.m.}$ = 15.69 and 18.63 MeV versus the angle in center-of-mass frame taken from F\"ul\"op \textsl{et al.} \cite{ful01}. The lines correspond to the predictions using different global optical potential parameter sets. For more information see Sec.~\ref{sec:global}.} 
\end{figure} 

Since modeling explosive nucleosynthesis scenarios requires reaction rates on
large number of reactions involving thousands of nuclei, the $\alpha$-nucleus
potential has to be known in a wide region. The reliability of the
extrapolation to unstable nuclei can be tested by measuring the elastic
scattering cross sections on several nuclei along isotopic and isotonic
chains. The ratio of Rutherford normalized cross sections along isotopic or
isotonic chains is a very sensitive observable for the quality of
$\alpha$-nucleus potentials that was not taken into account in most previous
studies. This interesting feature was not observed in earlier work because the
backward angular range was usually not measured with sufficient accuracy.

The $^{92}$Mo($\alpha,\alpha$)$^{92}$Mo reaction has been investigated by
F\"ul\"op \emph{et al.} at E$_{c.m.}$=13.20, 15.69 and 18.62 MeV \cite{ful01}.
For completeness, in Fig.~\ref{fig:mo} the angular distributions of
elastically scattered alpha particles on $^{92}$Mo measured at E$_{c.m.}$ =
15.69, 18.62 MeV are shown. The different lines correspond to theoretical
cross sections calculated from the above discussed global alpha nucleus
potentials, again without any further adjustment of parameters. As can be
seen, the situation is similar to the one found in the case of
$^{89}$Y($\alpha,\alpha$)$^{89}$Y. Namely, the Avrigeanu and McFadden global
parameterizations can describe the measured angular distributions with similar
quality and the potential of Kumar fails to reproduce the experimental data.

In order to investigate the behavior of the optical potential parameters along
the N = 50 isotonic chain we derived the ratio of the elastic scattering cross
sections of $^{92}$Mo and $^{89}$Y which are both neutron-magic nuclei. It is
found that the normalized elastic alpha scattering cross sections of $^{89}$Y
and $^{92}$Mo differ by roughly 50-70\% at backward angles, and the ratio
shows a pronounced oscillation pattern. The large number of experimental
points and the low uncertainties on both data sets provide a unique probe to
understand the evolution of the \al -nucleus potential along the $N = 50$
isotonic chain.

\begin{figure}
\resizebox{0.8\columnwidth}{!}{\rotatebox{270}{\includegraphics[clip=]{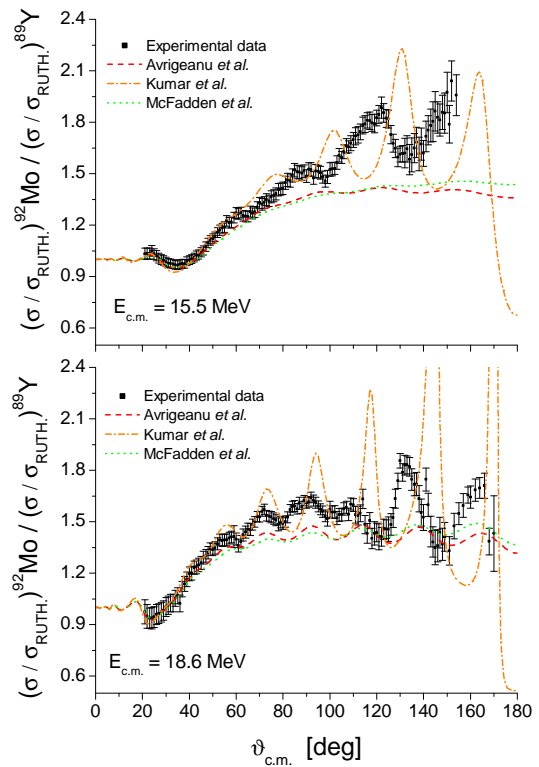}}}
\caption{
\label {fig:ratio} (color online).
Experimental ratio of the scattering cross sections ($\sigma/\sigma_{RUTH}$)$^{92}$Mo / ($\sigma/\sigma_{RUTH}$)$^{89}$Y at E$_{c.m.} \approx$ 15.51 and 18.63 MeV versus the angle in center-of-mass frame. The cross sections of the $^{92}$Mo($\alpha,\alpha$)$^{92}$Mo are taken from \cite{ful01}. The lines correspond to the predictions using different global optical potential parameter sets. For more information see Sec.~\ref{sec:global}.} 
\end{figure}

In Figure \ref{fig:ratio} the experimental ratio of the Rutherford normalized
elastic scattering cross sections is compared to the corresponding results of
the recent potential of Avrigeanu \textsl{et al.} \cite{avr07}, the most
recent global OMP from Kumar \emph{et al.} \cite{kum06} and the well-known
potential of McFadden and Satchler \cite{mcf}. It can be clearly seen that no
global parameterization can describe correctly the amplitude and the phase of
the oscillation pattern of the experimental data at backward angles. This new
observable may provide constraints for the further improvement of global \al
-nucleus potentials.

It may be added that, while the ROP \cite{avr07} reproduces only the phase of
oscillations rather well, the local Woods-Saxon parameter sets corresponding
to these target nuclei and incident energies (Table 2 of Ref. \cite{avr07})
provide a good description of this ratio at 15 MeV and a reasonable one at 19
MeV. Therefore, taking into account the ROP averaging nature, one may conclude
that the difference between the experimental ratio values and those provided
by the ROP describes the variance for particular nuclei with respect to the
average behavior.

Recently, a similar study has been performed by Galaviz \textsl{et
  al.}~\cite{gal05} where the variation of the elastic scattering cross
sections along the tin isotopic chain had been studied. Complete angular
distributions of the $^{112,124}$Sn($\alpha,\alpha$)$^{112,124}$Sn reactions
at 18.8 MeV were measured. It was found that the elastic alpha scattering
cross sections of the $^{112}$Sn and $^{124}$Sn differ by roughly 30-40\% at
backward angles, and the ratio of the measured cross sections shows a similar
oscillation feature. It became evident that the global alpha-nucleus
potentials failed to reproduce either the strength or the oscillation pattern
for backward angles \cite{gal05}. This behavior is very similar to the ratio
of the Rutherford normalized cross sections of the
$^{92}$Mo($\alpha,\alpha$)$^{92}$Mo and $^{89}$Y($\alpha,\alpha$)$^{89}$Y
derived in the present work. This fact clearly indicates that the available
theoretical alpha nucleus optical potential parameterizations have to be
improved to be able to describe the variation of the angular distributions
simultaneously along isotopic and isotonic chains. This is particularly
important for the extrapolation into regions of the chat of nuclides where no
scattering data exist.

\section{Summary}
\label{sec:sum}
In the present work angular distributions of elastically scattered alpha
particles on $^{89}$Y have been measured at E$_{c.m.}$ = 15.51 MeV and 18.63
MeV. The new experimental data have been used to determine the parameters of a
locally optimized folding potential. In order to investigate the energy
dependence of the potential parameters, angular distributions of
$^{89}$Y($\alpha$,$\alpha$)$^{89}$Y elastic scattering at higher energies
\cite{wit75, eng82, Bri72, Bin69, Als66} have also been analyzed. The
volume integrals of the local potential show a smooth variation with energy
over a broad energy range. However, the shape of the imaginary potential
cannot be strictly fixed at low energies. 

In addition, the local potential is also used to study bound state parameters
of the $^{93}$Nb = $^{89}$Y $\otimes$ \al\ system. Excitation energies of
excited states in $^{93}$Nb and their decay properties can be described
successfully with potential parameters very close to the scattering potential.

The new experimental scattering data have also been used as a sensitive test
for global parameterizations of the $\alpha$-nucleus potential that have to be
used in $p$ process network calculations.  Obviously, the best description of
the experimental data is obtained from the local potential where the
parameters are fitted to reproduce the measured angular
distributions. However, it is found that also the global parameterizations of
\cite{avr07} and \cite{mcf} provide a good description for the measured
angular distributions, and also reasonable agreement has been found for the
angular distributions at slightly higher energies that are available from
literature \cite{wit75,eng82}. For a deeper understanding of the differences
between the locally optimized potential and the global \al -nucleus
potentials, the calculated elastic phase shifts $\delta_L$ and scattered wave
amplitudes $\eta_L$ are presented.

Furthermore, the variation of elastic scattering angular distributions along
the $N = 50$ isotonic chain has been analyzed. Here all global
parameterizations failed to reproduce the amplitude and/or phase of the
oscillations of the ratio of the Rutherford normalized cross sections. In
order to advance our current understanding of the $\alpha$-nucleus optical
potential, further experimental scattering data with high precision are
essential as well as improvements of the available global \al -nucleus
potentials.

\begin{acknowledgments}
We would like to thank for A. Kumar for providing optical model calculations
by using the optical potential of \cite{kum06} for us.  This work was
supported by OTKA (K068801, T049245), by the European Research Council grant
agreement no. 203175 and DFG (SFB634 and ZI510/5-1).  G. G. K. and
D. G. acknowledge the support of the Spanish CICYT under the project
FPA2005-02379 and MEC Consolider project CSD2007-00042.  Gy. Gy\"urky
acknowledges support from the Bolyai grant. D. Galaviz is Juan de la Cierva
fellow (Spanish Ministry of Science).
\end{acknowledgments}

\end{document}